\documentclass[12pt]{iopart}
\usepackage{color}
\usepackage{ bbold }
\def\bea{\begin{eqnarray}}
\def\eea{\end{eqnarray}}

\begin{document}

\title[``Bethe-Ansatz-free" eigenstates of Richardson-Gaudin models]{``Bethe-Ansatz-free" eigenstates of spin-1/2 Richardson-Gaudin integrable models}
\author{Alexandre Faribault and Claude Dimo}
\address{ 
 Universit\'e de Lorraine, CNRS, LPCT, F-54000 Nancy, France }
 \ead{alexandre.faribault@univ-lorraine.fr}
\date{\today}

\begin{abstract}

In this work we construct the eigenstates of the most general spin-1/2 Richardson-Gaudin model integrable in an external magnetic field. This includes the possibility for fully anisotropic XYZ coupling such that the $S^x_iS^x_j$,   $S^y_iS^y_j$ and $S^z_iS^z_j$ terms all have distinct coupling strengths. While insuring that integrability is maintained in the presence of an external field excludes the elliptic XYZ model which is only integrable at zero field, this work still covers a wide class of fully anisotropic (XYZ) models associated with non skew-symmetric r-matrices.

The eigenstates, as constructed here, do not require any usable Bethe ansatz and therefore: no proper pseudo-vacuum, Bethe roots, or generalised spin raising (Gaudin) operators have to be defined. Indeed, the eigenstates are generically built only through the conserved charges which define the model of interest and the specification of the set of eigenvalues defining the particular eigenstate. Since these eigenvalues are, in general, solutions to a simple set of quadratic equations, the proposed approach is simpler to implement than any Bethe ansatz and, moreover, it remains completely identical independently of the symmetries of the model. Indeed, the construction removes any distinction between XYZ models and XXZ/XXX models and, generically, that between models with or without U(1) so that any difficulties associated with the use of a Bethe ansatz in any of these cases are avoided. 

\end{abstract}
\maketitle

\section{Introduction}

Quantum integrable models of the Richardson-Gaudin (RG) \cite{gaudin_1976,gaudin_bethe_2014,richardson_restricted_1963,richardson_exact_1964} class have already found applications in a remarkably vast array of domains: from superconductivity in metallic grains \cite{
cambiaggio_integrability_1997,dr-01,amico}, to nuclear pairing \cite{nuclear1,nuclear2,nuclear3} through applications in the behaviour of spin qubits coupled to a spin bath \cite{yuzbashyan_cs_2005,bortz_cs_2010,esebbag_elliptic_2015,seifert_cs_2016,claeys_cs_2018,frohling_cs_2018}. Their value to problems of physical interest cannot be denied and, as such, the capacity to compute their physical properties as efficiently as possible remains an important question.

In recent years, this class of integrable models has been considerably expanded by relaxing the usual assumption of anti-symmetric couplings. Either through non-skew-symmetric r-matrices  \cite{skrypnyk_generalized_2007,skrypnyk_non-skew-symmetric_2009,skrypnyk_non-skew-symmetric_2012,skrypnyk_elliptic_2012,skrypnyk_twisted_2016,antonio_boundary_2013,lukyanenko_boundaries_2014,lukyanenko_integrable_2016,links_solution_2017} or, as in Gaudin's original work \cite{gaudin_1976,gaudin_bethe_2014} by directly building the commuting conserved charges  \cite{claeys_readgreen_2016,faribault_field_2017}.  It was demonstrated that a wide class of such models remain integrable in the presence of an arbitrary external magnetic field, not only in the XXX and XXZ models, but also for certain fully anisotropic XYZ RG models built out of spins-1/2. While similar constructions are possible for higher spins $s>1/2$ (see \cite{skrypnyk_elliptic_2012} for example), they must contain additional local terms which takes them outside of the class of RG models studied here which is assumed to only contain $S^x_iS^x_j$,   $S^y_iS^y_j$ and $S^z_iS^z_j$ coupling terms between spins $i$ and $j$.

It was recently demonstrated that, for these spin-1/2 systems in an external field, the conserved charges systematically obey quadratic relations so that the eigenvalues characterising their eigenstates can always be found as the solutions to a small set of quadratic equations \cite{dimo_quadratic_2018}. Expectation values of local spin operators, in any of these eigenstates, can then also be computed easily by making use of the Hellmann-Feynman theorem giving access to some of the physics of these systems \cite{claeys_xyz_2018}. All of these results can be obtained without any knowledge of whether, or how, one can construct the eigenstates corresponding to these eigenvalues.

When the system of interest retains a U(1) symmetry (associated to its rotational invariance in a given plane we will call the XY plane), they allow for the use of the Algebraic Bethe Ansatz (ABA) to construct their eigenstates \cite{korepin_book,zhou_aba_2002,links_aba_2003,ortiz_exactly-solvable_2005}. In doing so, one defines the eigenvalues in terms of a set of Bethe roots $\{\lambda_1 \dots \lambda_M\}$. These Bethe roots, when ``on-shell", i.e. solutions of a system of $M<N$ coupled non-linear algebraic equations dubbed the Bethe equations, provide the various eigenvalues of the conserved charges associated with the corresponding eigenstate. This set of Bethe roots then also provide, through the repeated action $M$ times of a generalised creation operator $B(u)$ a representation of the eigenstate itself as $\prod_{i=1}^M B(\lambda_i) \left| \Omega\right>$, where the specific state $\left| \Omega\right>$ can be called the pseudo-vacuum.

Certain models which break $U(1)$ symmetry either through an external magnetic field \cite{lukyanenko_integrable_2016,claeys_readgreen_2016,faribault_field_2017}, or through other integrable ``boundaries" \cite{lukyanenko_boundaries_2014,crampe_MABA_2017} which can lead to additional terms in the conserved charges, have also be shown to be solvable in a remarkably similar way despite lacking a ``proper" pseudo-vacuum. A construction similar to the ABA has indeed been shown to lead to eigenstates of the form $\prod_{i=1}^N B(\lambda_i) \left| \Omega\right>$ (now containing a number of Bethe roots equal to $N$, the system size) acting on the same reference-state as the one used for a normal ABA implementation. This Modified Algebraic Bethe Ansatz (MABA) \cite{crampe_MABA_2017,belliard_MABAXXX_2013,belliard_MABA_2015,belliard_MABA2_2015,avan_MABA_2015} therefore also provides a compact representation of the eigenstates of these models. However, for XYZ models, which break $U(1)$ symmetry by being fully anisotropic in their couplings, only the elliptic Gaudin models has, to the best of our knowledge, an explicit construction of its eigenstates and it requires a much more complex ABA \cite{sklyanin_xyz_1996}. The XYZ RG models in an external magnetic field which are discussed in this work do not naturally give access to an ABA or MABA approach for finding their eigenstates. Nonetheless, explicit expressions for the eigenstates will be built, only by exploiting the quadratic relations between the spin-1/2 conserved charges and, therefore, without having to define any Bethe ansatz procedure. Since the approach completely bypasses any reference to Bethe roots, to the need to specify a pseudo-vacuum state or to the necessity to define proper $B(u)$ operators, it provides what can be dubbed a ``Bethe-ansatz-free" solution of the generic (XYZ) spin-$\frac1{2}$ Richardson-Gaudin integrable models in an arbitrary magnetic field.

The resulting construction gives explicit expression for the projectors on any eigenstate in terms of only the conserved charges of the model and the set of eigenvalues defining the eigenstate of interest. Since finding these eigenvalues has been reduced to finding solutions of a system of quadratic equations, the resulting formal definition of eigenstates could become of great use numerically, since it only requires the easy to compute (compared to Bethe roots) eigenvalues.

The next section will first describe the models this work is concerned with as well as the relevant quadratic relations between conserved charges and eigenvalues. In the following one, the main result of this work is given, namely an explicit expression for the eigenstate projectors as operator-determinant, after which the validity of the construction is proven and a simple resolution of the identity operator is proposed. Finally, we present an explicit expansion of the determinant which highlights some of its underlying mathematical structure while concluding remarks are given in the very last section.

\section{The models}

The quantum integrable models of interest in this work are all characterised by the existence of $N$ operators $\hat{R}_i$ commuting with each other. Here $N$ is the number of spins (or pseudo-spins) in the system and each of them will be considered as a spin-1/2 representation of an SU(2) algebra, i.e. each of them (labelled by an index $i=1, 2 \dots N$) are defined by the usual $2\times2$ Pauli matrices  $\sigma^x_i,\sigma^y_i, \sigma^z_i$. 

We impose that the conserved charges have only diagonal-couplings ($x$ direction coupled with $x$ direction, etc) and allow for an external magnetic field, i.e. the conserved charges are chosen to have the form:
\bea
\hat{R}_i = \vec{B}_i \cdot \vec{\sigma}_i + \sum_{j\ne i}^N \left(X_{ij} \sigma^x_i\sigma^x_j + Y_{ij} \sigma^y_i\sigma^y_j+Z_{ij} \sigma^z_i\sigma^z_j\right).
\eea

\noindent One can then insure integrability by simply making sure that the parameters are chosen in such a way that $\left[ \hat{R}_i,\hat{R}_j\right] = 0 \ \ \forall \ \ i,j$. This imposes restrictions on the coupling constants $X_{ij},Y_{ij},Z_{ij}$ and the field components $B^x_i,B^y_i, B^z_i$. The solutions to these constraints have been completely parametrised in \cite{claeys_xyz_2018} where it was demonstrated that the most general set of parameters compatible with the requirements of integrability can be written as:

\begin{eqnarray}
B^x_i = \frac{\gamma}{F_x(\epsilon_i)},&\ \ \ &X_{ij} = g\frac{F_x(\epsilon_i)F_y(\epsilon_j)}{\epsilon_i-\epsilon_j} \nonumber\\
B^y_i = \frac{\lambda}{F_y(\epsilon_i)},  &\  \ \ &Y_{ij} =  g\frac{F_x(\epsilon_j)F_y(\epsilon_i)}{\epsilon_i-\epsilon_j}\nonumber\\
B^z_i = 1, &\  \ \ &Z_{ij} = g\frac{F_x(\epsilon_j)F_y(\epsilon_j)}{\epsilon_i-\epsilon_j},
\label{genparam} 
\end{eqnarray}

\noindent with the functions $F$ given by:  

\bea
F_x(\epsilon_j) &=&\sqrt{\alpha_x \epsilon_j + \beta_x}
\nonumber\\
F_y(\epsilon_j)&=& \sqrt{\alpha_y \epsilon_j + \beta_y},
\eea

\noindent for arbitrary $g,\gamma,\lambda,\alpha_x,\alpha_y, \beta_x, \beta_y$ and $\epsilon_i$. Here, the choice of $B^z_i = 1$ remains generic in that conserved charges can always be rescaled by an arbitrary constant. Since $\frac{\hat{R}_i}{B^z_i}$ still all commute together, any model with $N$ arbitrary non-zero values $B^z_i$ can always be rescaled to an equivalent $B^z_i =1$ model. The only case which is therefore excluded is that of models which are integrable exclusively at strictly zero field, namely the elliptic Gaudin models. Indeed, although zero-field limits of the models considered here can be properly defined through an appropriate rescaling of the parameters \cite{claeys_xyz_2018}, in the fully anisotropic XYZ case such a limit does not correspond to the usual elliptic Gaudin models. The elliptic models being antisymmetric in their coupling constants while the zero-field limit of the XYZ models of interest here can only lead to non-skew-symmetric XYZ models.

For all of the models parametrised by (\ref{genparam}), it was shown \cite{dimo_quadratic_2018} that integrability is sufficient to establish that the resulting $N$ conserved charges obey, at the operator level, the following quadratic relations:
\bea
\hat{R}_i ^2 = \sum_{j\ne i}^N \Gamma_{ij} \hat{R}_j + K_i,
\label{quadoperator}
\eea 

\noindent where $\Gamma_{ij} = - \frac{2 Z_{ij} Y_{ij}}{X_{ji}}$ and $K_{i} = \sum_{\alpha} B_\alpha^2+  \sum_{j\ne i} \left(X_{ij}^2+Y_{ij}^2+Z_{ij}^2\right)$. Since the general parametrisation, once rescaled to $B^z_i=1$, systematically leads to $X_{ij} = -Y_{ji}$ one can then always use these adequately rescaled operators and therefore set:
\bea
\Gamma_{ij} = 2 Z_{ij}. 
\eea

Since they all commute with one another, the conserved charges share a common eigenbasis in which they are all diagonal and therefore the same quadratic relations are also obeyed by the $N$ eigenvalues $(r_1\dots r_N)$ which specify any given eigenstate. The $n^{\mathrm{th}}$ eigenstate: $\left| \psi_n\right>$ such that $\hat{R}_i \left|\psi_n\right> = r_i^n\left|\psi_n\right> \ \forall \ i=1 \dots N$ is then, in principle, fully characterised by $(r_1^n \dots r_N^n)$, i.e. the $n^{\mathrm{th}}$ solution to the system of quadratic equations:
\bea
r_i ^2 = \sum_{j\ne i}^N \Gamma_{ij} r_j + K_i.
\label{quadeigen}
\eea

In the next section, we demonstrate that the projector $P_n \equiv \left| \psi_n \right>\left< \psi_n \right|$, on any given eigenstate $\left| \psi_n \right>$ can be defined in a compact way, exclusively in terms of the conserved charges $\hat{R}_i$ and the corresponding set of eigenvalues $r^n_i$ of that particular eigenstate. The resulting representation of the eigenstates is then obtainable exclusively and directly from the solutions of the quadratic Bethe equations for the eigenvalues (\ref{quadeigen}) whose simplicity provides an important numerical advantage in stability and computational speed \cite{faribault_xxxsolve_2011,elaraby_xxxsolve_2012,claeys_xxzsolve_2015,faribault_cs_2013,faribault_cs2_2013} over the (M)ABA approaches based on Bethe roots. Even in the XXZ cases (including those without U(1)-symmetry), for which eigenstates could be written in terms of Bethe roots, the proposed representation could lead to major simplification in numerically approaching the physics of these systems. 

\section{Eigenstate projectors}

A direct consequence of the quadratic relations (\ref{quadoperator}) between operators is that an arbitrary polynomial (or even a formal power series) in these conserved charges: $  \hat{Q}(\hat{R}_1 \dots \hat{R}_N) \equiv \displaystyle\sum_{n_1,n_2\dots n_N=0}^\infty C_{n_1 \dots n_N} \ \hat{R}_1^{n_1}\hat{R}_2^{n_2} \dots \hat{R}_N^{n_N}$, can always be reduced to:
\bea
\hat{Q}(\hat{R}_1 \dots \hat{R}_N) = \sum_{n_1,n_2 \dots n_N=0}^1 B_{n_1 \dots n_N} \ \hat{R}_1^{n_1}\hat{R}_2^{n_2} \dots \hat{R}_N^{n_N};
\label{linearpoly}
\eea 

\noindent a polynomial which is at most linear in each of the conserved charges. As a simple example for $N=2$, one has $\hat{R}_1^3\hat{R}_2 = \hat{R}_1 (\Gamma_{12}\hat{R}_2 +K_1) \hat{R}_2 = \Gamma_{12}\hat{R}_1 \hat{R}^2_2 + K_1\hat{R}_1  \hat{R}_2 =  \Gamma_{12}\hat{R}_1 (\Gamma_{21} \hat{R}_2 + K_2)+ K_1\hat{R}_1  \hat{R}_2  =    (\Gamma_{12}\Gamma_{21}  + K_1) \hat{R}_1\hat{R}_2 + K_2 \Gamma_{12}\hat{R}_1$. Through these successive reduction of quadratic terms to linear ones, the same procedure could be carried out for arbitrary orders and for arbitrary number of conserved charges $N$, systematically leading to polynomials at most linear in each $\hat{R}_i$. Such a generic linear polynomial is then, ultimately, defined by a set of $2^N$ coefficients $B_{n_1 \dots n_N}$, a number which corresponds to the dimension of the full Hilbert space.

Among this large class of operators parametrised by $2^N$ coefficients, we define a subclass, compactly parametrised by only $N$ real variables $(r_1 \dots r_N)$ as the following determinant of an ``operator-valued" $N$ by $N$ matrix:

\bea
\hat{P}(r_1 \dots r_N) \equiv \mathrm{det} \hat{J} \ \ \ \mathrm{ with } 
\left\{\begin{array}{l} 
\hat{J}_{ii} =  r_i+\hat{R}_i \\
\hat{J}_{ij} =   -\Gamma_{ij} \ \ \ \forall \ i \ne j.
\end{array}
\right. 
\label{opdet}
\eea

\noindent Keeping in mind that every conserved charges commute with one another, the determinant representation of $\hat{P}$, despite having operator entries, has no ambiguity in the order of the products involved. We define as well the following scalar-valued polynomial:
\bea
N(r_1 \dots r_N) \equiv \mathrm{det} N \ \ \ \mathrm{ with } 
\left\{\begin{array}{l} 
N_{ii} =  2r_i \\
N_{ij} =   -\Gamma_{ij} \ \ \ \forall \ i \ne j.
\end{array}
\right.
\label{scaldet}
\eea
In the next section, we will prove our main result which is stated here: the projector on the normalised eigenstate $\left|\psi_{n}\right>$ such that $\hat{R}_i \left|\psi_{n}\right> = r_i^n\left|\psi_{n}\right>$, is proportional to the operator $\hat{P}$ evaluated at $(r^n_1 \dots r^n_N)$:

\bea
&& \left|\psi_n\right>\left<\psi_n\right|  \propto \hat{P}(r_1^n \dots r_N^n) \nonumber\\ && = \mathrm{det}\left(\begin{array}{ccccc} 
r_1^n + \hat{R}_1 & -\Gamma_{12} & -\Gamma_{13} & \dots & -\Gamma_{1N} \\
-\Gamma_{21} & r_2^n + \hat{R}_2& -\Gamma_{23} & \dots & -\Gamma_{2N} \\
-\Gamma_{31} & -\Gamma_{32} & r_3^n + \hat{R}_3 & \dots & -\Gamma_{3N} \\
\vdots &\vdots &\vdots &\ddots &\vdots \\
-\Gamma_{N1} & -\Gamma_{N2} & -\Gamma_{N3} & \dots & r_N^n + \hat{R}_N \\
\end{array}
\right)
\nonumber\\
\label{projector}
\eea

\noindent and can be properly normalised by the scalar determinant defined in (\ref{scaldet}) so that:
\bea
\left|\psi_n\right>\left<\psi_n\right| = \frac{\hat{P}(r^n_1 \dots r^n_N)}{N(r^n_1 \dots r^n_N)}.
\label{normalised}
\eea

Consequently, an (unnormalised) representation of the eigenstate itself can be obtained from the action of this projector on an arbitrary ``vacuum" state $\left|\Omega\right>$ as:
\bea
\left| \psi_n \right> \propto \hat{P}(r_1^n \dots r_N^n) \left| \Omega \right>,
\label{eigenstate}
\eea
\noindent with the only requirement that $\left| \Omega \right>$ has a non-zero overlap with the target eigenstate $\left| \psi_n \right>$. This is not particularly limiting since, in a generic model, eigenstates will have a non-zero overlap with any canonical spin state  $\prod_{i=1}^r S^+_{i_r} \left|\downarrow \downarrow \dots \downarrow\right>$ which could then all serve the role of vacuum. The exception is for models which exclusively contains a z-component of the magnetic field. In these cases, the XXZ (and XXX) models retain the U(1) symmetry and therefore each eigenstate have a fixed z-axis total magnetisation. This would require $\left| \Omega \right>$ to have a component in the appropriate magnetisation sector. Consequently, any $\left| \Omega \right>$ which is spread over every sector would provide a proper ``vacuum" for every eigenstate. For XYZ models in a z-only magnetic field, the eigenstates actually have fixed magnetisation parity, meaning they either belong to the sub-Hilbert-space which contains only odd OR even numbers of up-pointing spins. Consequently any  $\left| \Omega \right>$ which contains at least one odd and one even magnetisation term (such as $\left| \downarrow \downarrow \dots \downarrow \right> + \left| \uparrow \downarrow \dots \downarrow \right> $ for example) would allow one to build any eigenvector. 

While the projectors themselves (\ref{normalised}) are indeed the ones built using the normalised eigenstate, the resulting vector (\ref{eigenstate}) would, however, have a normalisation which depends on the particular choice of $\left|\Omega\right>$.

\section{Proof of the projector's representation}

Considering that the family of operators $\hat{P}(r_1 \dots r_N)$ as defined by eq. (\ref{projector}) are built exclusively out of the conserved charges and scalar parameters, $\hat{P}(r_1 \dots r_N)$ is always a diagonal operator in the orthonormalised eigenbasis built from the $\left| \psi_n \right>$ eigenstates. Each one of the matrix elements in the determinant (\ref{projector}) is therefore also a diagonal matrix in this basis. The orthogonality of the basis states consequently leads to the following explicit expansion of the determinant in that basis:
\bea
\hat{P}(r_1 \dots r_N)  = \sum_{m=1}^{2^N}  C_m(r_1 \dots r_N)\left| \psi_m\right> \left<\psi_m\right|,
\label{diagonalexpansion}
\eea
\noindent where
\bea
&&C_m(r_1 \dots r_N) = \mathrm{det}\left(\begin{array}{ccccc} 
r_1 + r_1^m  & -\Gamma_{12} & -\Gamma_{13} & \dots & -\Gamma_{1N} \\
-\Gamma_{21} & r_2 + r_2^m & -\Gamma_{23} & \dots & -\Gamma_{2N} \\
-\Gamma_{31} & -\Gamma_{32} & r_3 + r_3^m & \dots & -\Gamma_{3N} \\
\vdots &\vdots &\vdots &\ddots &\vdots \\
-\Gamma_{N1} & -\Gamma_{N2} & -\Gamma_{N3} & \dots & r_N + r_N^m  \\
\end{array}
\right).
\nonumber\\
\label{cncoeff} 
\eea

In order to prove that $P(r_1 \dots r_N)$ becomes proportional to the desired projector when evaluated at $(r_1^n \dots r_N^n)$, it is sufficient to prove that, at this point, every coefficient $C_{m \ne n}(r^n_1 \dots r^n_N)$ cancels out.

Considering that, being eigenvalues $(r_1^n \dots r_N^n)$ and $(r_1^m \dots r_N^m)$ are ``on-shell", i.e. they form a solution to the quadratic Bethe equations (\ref{quadeigen}), one has:
\bea
(r^n_i)^2 - (r_i^m )^2 = \sum_{j \ne i}^N \Gamma_{ij} \left(r^n_{j} - r^m_{j}\right),
\eea

\noindent it then becomes easy to see that, when the matrix $J$ given in eq. (\ref{cncoeff}) is evaluated at  $(r_1 \dots r_N)=(r_1^n \dots r_N^n)$, its columns are not independent since the following linear combination of its columns:

\bea
\sum_{j=1}^N J_{ij} (r_j^n-r_j^m) = (r_i^n+r_i^m)(r_i^n-r_i^m) - \sum_{j\ne i}^N \Gamma_{ij} \left(r^n_{j} - r^m_{j}\right)=0. 
\eea

Consequently, the determinant, and therefore $C_m(r^n_1 \dots r^n_N)=0$ for every $n\ne m$. The only exception is found when $n=m$ since every coefficient in the linear combination is then explicitly $(r_j^n-r_j^n) =0$. In this case, the non-zero value of $C_n(r^n_1 \dots r^n_N)$ is obviously given by the determinant polynomial $N(r_1^n \dots r_N^n)$ defined in (\ref{scaldet}). This therefore proves the proposition: the operator $\hat{P}(r_1 \dots r_N)$ introduced in eq. (\ref{opdet}), when evaluated at a point $(r_1 \dots r_N)$ which corresponds to a solution of the quadratic Bethe equations (\ref{quadeigen}), become the projector on the normalised eigenstate corresponding to these eigenvalues:

\bea
 \left| \psi_n\right>\left<\psi_n\right| = \frac{\hat{P}(r^n_1 \dots r^n_N)}{N(r^n_1 \dots r^n_N)} .
\eea

This provides an operator-based concept of ``on-shell" and ``off-shell", where the generic $N$-parameter operator $\hat{P}(r_1 \dots r_N)$, becomes the projector on an eigenstate of the integrable model whenever its $N$ free parameters are on-shell, i.e. chosen to be a solution of (quadratic) Bethe equations. This is in complete analogy with the usual ABA for which a generic Bethe state $\prod_{i=1}^M B(\lambda_i)\left|\Omega\right>$ becomes an eigenstate whenever the Bethe roots $\lambda_i$ are  on-shell, i.e.: solution to the Bethe equations. However, the construction proposed in this work avoids the need to define the relation between Bethe roots and eigenvalues of the conserved charges, the need to define the right $B(u)$ operators and the necessity to find a proper pseudo-vacuum $\left|\Omega\right>$ on which to create the excitations. In this sense, it gets completely rid of the possible difficulties of applying a Bethe ansatz and provides, nonetheless, the exact solution to the eigenproblem of spin-1/2 Richardson-Gaudin integrable systems even in the cases where Bethe roots-based approaches typically cannot be used. The proposed solution remains valid in every case without U(1) symmetry (apart from the zero-field elliptic model which does not belong to the class of models for which the proposed quadratic equations are respected). Among these, the XXZ  (and XXX) systems in an arbitrary external field could also be approached using the (M)ABA, but the XYZ case in an arbitrary field did not, to the best of our knowledge, have an explicit solution for its eigenstates yet.

\subsection{Off-shell resolution of the identity and product of operators}

On top of providing, when on-shell, a representation for the eigenstate projectors, the generic (off-shell) family of operators $\hat{P}(r_1 \dots r_N)$ also allows one to build simple resolutions of the identity. Indeed, as it was defined in (\ref{opdet}), $\hat{P}(r_1 \dots r_N)$ is not only a polynomial in the conserved charges $\hat{R}_i$, but also a polynomial, of maximal order 1, in each of the scalar variables $r_i$. Moreover, the (operator) coefficient  in front of the $\left(\displaystyle\prod_{i=1}^N r_i\right)$ term is given by $\mathbb{1}$ since this term, as seen by expanding the determinant, comes exclusively from the product of the $N$ diagonal elements. Consequently, integrating $\Psi(r_1 \dots r_N) \hat{P}(r_1 \dots r_N)$ on an N-dimensional hypercube centered around the origin will provide, for any function $\Psi(r_1 \dots r_N)$ antisymmetric in every variable, a resolution of the identity. A simple example would be to choose $\Psi(r_1 \dots r_N) = \prod_{i=1}^N r_i$ and integrate, on the real axis, on the symmetric intervals $r_i \in \left[-a,a\right]$. This leads to
\bea
\int_{-a}^{a}dr_1 \dots \int_{-a}^{a} dr_N \left(\prod_{i=1}^N r_i\right) \hat{P}(r_1 \dots r_N) = \left(\frac{2a^3}{3}\right)^N \mathbb{1},
\nonumber\\
\label{identity}
\eea
\noindent since every term in the expansion of $\hat{P}(r_1 \dots r_N)$, except for the leading term $\left(\prod_{i} r_i\right)$, is missing at least one of the $r_i$ variables. Apart from the leading term, every term in $\Psi(r_1 \dots r_N) \hat{P}(r_1 \dots r_N)$ is odd in the missing variables and their integral over the symmetric intervals will therefore cancel out. 

Another important quantity which is simple to evaluate is the product of an ``on-shell" and ``off-shell" operator, analogous to Slavnov's determinants \cite{zhou_aba_2002, links_aba_2003, slavnov_1989,claeys_inner_2017}. In fact, considering that we have an explicit decomposition of the ``off-shell" operator $\hat{P}(r_1 \dots r_N)$ given in (\ref{diagonalexpansion}) we directly know that its product with an ``on-shell" projector is given by the corresponding coefficient (\ref{cncoeff}), namely:
\bea
&&\hat{P}(r_1 \dots r_N)\hat{P}(r^n_1 \dots r^n_N) \nonumber\\ &&= \mathrm{det}\left(\begin{array}{ccccc} 
(r_1 + r_1^n ) & -\Gamma_{12} & -\Gamma_{13} & \dots & -\Gamma_{1N} \\
-\Gamma_{21} & (r_2 + r_2^n )& -\Gamma_{23} & \dots & -\Gamma_{2N} \\
-\Gamma_{31} & -\Gamma_{32} & (r_3 + r_3^n ) & \dots & -\Gamma_{3N} \\
\vdots &\vdots &\vdots &\ddots &\vdots \\
-\Gamma_{N1} & -\Gamma_{N2} & -\Gamma_{N3} & \dots & (r_N + r_N^n ) \\
\end{array}
\right) \left|\psi_n\right>\left< \psi_n\right|.
\eea
This result is highly reminiscent of the eigenvalue-based determinants known for XXX \cite{faribault_field_2017,faribault_det_2012} and XXZ \cite{claeys_xxzsolve_2015,claeys_inner_2017} models, being again defined simply as determinants in which eigenvalues $r^n_i$ and their ``off-shell" equivalent $r_i$ only appear as their sum $(r_i+r^n_i)$ and do so exclusively in the diagonal elements of an $N$ by $N$ matrix.
 
\section{Explicit expansion of $\hat{P}(r_1 \dots r_N)$}

In this last section, we study the explicit expansion of the family (\ref{opdet}) of operators $\hat{P}(r_1 \dots r_n)$ in a way which provides further insight into their underlying structure. Considering the general expression of a determinant as a sum over permutations:

\bea
\hat{P}(r_1 \dots r_N) = \sum_{\sigma \in S_n} \mathrm{sgn}(\sigma) \prod_{i=1}^n \hat{J}_{i, \sigma_i},
\eea

\noindent the first term present is given by the ``correctly" ordered permutation $\sigma_1 = 1, \sigma_2 =2 \dots \sigma_N=N$ and leads, in the determinant (\ref{opdet}), to the term $\prod_{i=1}^N (r_i + R_i)$. Permutations which exchange only two of these indices $(\sigma_{k'}=k,\sigma_k=k')$ would then lead to the replacement of two diagonal elements by the corresponding off-diagonal elements, while changing the sign of the permutation. These then give a series of terms $\displaystyle - \sum_{k \ne k'}^N \left[ \Gamma_{k k'}\Gamma_{k' k}\prod_{i\ne \{k,k'\}}^N (r_i + R_i)\right]$, etc. Since the determinant is a sum over products of $N$ matrix elements $\hat{J}_{ij}$ where the $i$ indices need to cover every line and the $j$ indices have to cover every column, one can classify the resulting terms according to the number of diagonal matrix elements (when $\sigma_k = k$) which are used to build each individual term. One can then write the determinant as a sum over the subsets $\pi$ of every possible cardinality $r = 0,1 \dots N$, that one can build out of $\{1,2,3 \dots N\}$.  A given set  $\pi = \{\pi_1, \pi_2 \dots \pi_r\}$ defines the diagonal matrix elements $\hat{J}_{\pi{_k}\pi{_k}}$ present and the complement $\bar{\pi} = \{\bar{\pi}_1 \dots \bar{\pi}_{N-r} \}$ (such that $\pi \cup \bar{\pi} = \{1,2,3 \dots N\}$) defines the indices which will be used for off-diagonal elements.  Defining $\Pi$ as the power set of $\{1,2,3 \dots N\}$ (containing every one of its subsets of every cardinality), one can write:

\bea
\hat{P}(r_1 \dots r_N) = \sum_{\pi \in \Pi} K_\pi \prod_{k=1}^r(r_{\pi_k} +R_{\pi_k})
\eea
\noindent with $K_\pi$ given by the $M \times M$ determinant (with $M \equiv N-r$):

\bea
 K_\pi= \mathrm{det}\left(\begin{array}{ccccc} 
0 & -\Gamma_{\bar{\pi}_1\bar{\pi}_2 } & -\Gamma_{\bar{\pi}_1\bar{\pi}_3} & \dots & -\Gamma_{\bar{\pi}_1\bar{\pi}_{M}} \\
-\Gamma_{\bar{\pi}_2\bar{\pi}_1} & 0& -\Gamma_{\bar{\pi}_2\bar{\pi}_3} & \dots & -\Gamma_{\bar{\pi}_2\bar{\pi}_{M}} \\
-\Gamma_{\bar{\pi}_3\bar{\pi}_1} &- \Gamma_{\bar{\pi}_3\bar{\pi}_2} &0 & \dots & -\Gamma_{\bar{\pi}_3\bar{\pi}_{M}} \\
\vdots &\vdots &\vdots &\ddots &\vdots \\
-\Gamma_{\bar{\pi}_{M}\bar{\pi}_1} & -\Gamma_{\bar{\pi}_{M}\bar{\pi}_2} & -\Gamma_{\bar{\pi}_{M}\bar{\pi}_3} & \dots & 0 \\
\end{array}
\right),
\nonumber\\
\eea

\noindent built from the $M$ indices in the complement $\{\bar{\pi}_1 \dots \bar{\pi}_{M}\}$. Remarkably, the structure of $\Gamma_{i,j}$ imposed by integrability is such that, for $M$ odd, the resulting coefficient is strictly zero. Indeed, using the general parametrisation (\ref{genparam}) and the fact that $\Gamma_{ij} = 2 Z(\epsilon_i,\epsilon_j)$, one can write the determinant as:

\bea
&& K_\pi = \left(\displaystyle \prod_{ k=1}^{ M} -2g\ F_x(\epsilon_{\bar{\pi}_k}) F_y(\epsilon_{\bar{\pi}_k})\right)  \nonumber\\ && \times\mathrm{det}\left(\begin{array}{ccccc} 
0 & \frac{1}{\epsilon_{\bar{\pi}_1} -\epsilon_{\bar{\pi}_2} }& \frac{1}{\epsilon_{\bar{\pi}_1} -\epsilon_{\bar{\pi}_3} } & \dots & \frac{1}{\epsilon_{\bar{\pi}_1} -\epsilon_{\bar{\pi}_{ M}} } \\
\frac{1}{\epsilon_{\bar{\pi}_2} -\epsilon_{\bar{\pi}_1} } & 0&\frac{1}{\epsilon_{\bar{\pi}_2} -\epsilon_{\bar{\pi}_3} } & \dots &  \frac{1}{\epsilon_{\bar{\pi}_2} -\epsilon_{\bar{\pi}_{ M}}} \\
\frac{1}{\epsilon_{\bar{\pi}_3} -\epsilon_{\bar{\pi}_1} }& \frac{1}{\epsilon_{\bar{\pi}_3} -\epsilon_{\bar{\pi}_2} }&0 & \dots &  \frac{1}{\epsilon_{\bar{\pi}_3} -\epsilon_{\bar{\pi}_{ M}}}  \\
\vdots &\vdots &\vdots &\ddots &\vdots \\
\frac{1}{\epsilon_{\bar{\pi}_{ M}} -\epsilon_{\bar{\pi}_1} } & \frac{1}{\epsilon_{\bar{\pi}_{ M}} -\epsilon_{\bar{\pi}_2} }&\frac{1}{\epsilon_{\bar{\pi}_{ M}} -\epsilon_{\bar{\pi}_3} }& \dots & 0 \\
\end{array}
\right),
\nonumber\\
\eea

\noindent showing it is proportional to that of a square real skew-symmetric matrix $A$ of dimension $N-r$. From Jacobi's theorem, we then know that the resulting determinant is zero for any odd $N-r$ dimension. For even dimensions, ``Cayley's theorem for pfaffians" states that the determinant of a skew-symmetric matrix is given by the square of its pfaffian. However, in the case at hand, the integrability of the models, and the constraints it imposes on the matrix elements $\Gamma_{ij}$, the non-zero determinants found for even $N-r$ can be further simplified into a single sum over partitions into pairs. Redefining, for compactness, and without loss of generality, the indices $\bar{\pi}_i \to i $, we are interested in the determinant of the matrix $A_{i,j} = \frac{1-\delta_{ij}}{\epsilon_i - \epsilon_j}$.

By the very definition of a determinant, one has to encounter twice the index 1 in each term of its expansion, one matrix element having to come from the first line and one from the first column. Looking at a given specific pairing, say $(\epsilon_1, \epsilon_2)$, the expansion of the determinant can therefore contain: a) a term quadratic in $\frac{1}{(\epsilon_1-\epsilon_2)}$ coming from $A_{12}A_{21}$, b) terms linear in $\frac{1}{(\epsilon_1-\epsilon_2)}$ which comes from either $A_{12}$ or $A_{21}$ while the other instance of index 1 comes from a different line or column than $2$ and finally c) terms which do not contain $\frac{1}{(\epsilon_1-\epsilon_2)}$, i.e. terms which do not contain $A_{12}$ or $A_{21}$. 

Such a classification of terms allows one to study the analytic structure of the determinant by looking at $\epsilon_1 = \epsilon_2 + \Delta$ in the limit $\Delta \to 0$ in order to prove that the determinant has no simple pole at $\epsilon_1 \to \epsilon_2$ and therefore that terms linear in $\frac{1}{(\epsilon_1-\epsilon_2)}$ all sum up to zero.

Expanding the determinant in its first two rows, it can be written as:

\bea
&&\lim_{\Delta \to 0} \mathrm{det} A
=
\mathrm{det}\left( \begin{array}{cccccc} 
0 & -\frac{1}{\Delta} & \frac{1}{\epsilon_{  2} -\epsilon_{  3} }& \frac{1}{\epsilon_{  2} -\epsilon_{  4} } & \dots & \frac{1}{\epsilon_{  2} -\epsilon_{  { M}} } \\
\frac{1}{\Delta}  & 0&\frac{1}{\epsilon_{  2} -\epsilon_{  3} } & \frac{1}{\epsilon_{  2} -\epsilon_{  4} }& \dots &  \frac{1}{\epsilon_{  2} -\epsilon_{  { M}}} \\
\frac{1}{\epsilon_{  3} -\epsilon_{  2} }& \frac{1}{\epsilon_{  3} -\epsilon_{  2} }&0 & \frac{1}{\epsilon_{  3} -\epsilon_{  4} }& \dots &  \frac{1}{\epsilon_{  3} -\epsilon_{  { M}}}  \\
\frac{1}{\epsilon_{  4} -\epsilon_{  2} }& \frac{1}{\epsilon_{  4} -\epsilon_{  2} } & \frac{1}{\epsilon_{  4} -\epsilon_{  3} }&0& \dots &  \frac{1}{\epsilon_{  4} -\epsilon_{  { M}}}  \\
\vdots &\vdots &\vdots&\vdots &\ddots &\vdots \\
\frac{1}{\epsilon_{  { M}} -\epsilon_{  2} } & \frac{1}{\epsilon_{  { M}} -\epsilon_{  2} }&\frac{1}{\epsilon_{  { M}} -\epsilon_{  3} }&\frac{1}{\epsilon_{  { M}} -\epsilon_{  4} } & \dots & 0 \\
\end{array}
\right)
\nonumber\\ &=&
\mathrm{det}\left(\begin{array}{cc} 
0 & -\frac{1}{\Delta} \\
\frac{1}{\Delta} & 0 
\end{array}
\right)
\mathrm{det}{A^{\hat{1},\hat{2}}_{\hat{1},\hat{2}}}
+  \sum_{j=3}^{N-r}  \epsilon^{1j} \ \mathrm{det}\left(\begin{array}{cc} 
0 & \frac1{\epsilon_2-\epsilon_j} \\
\frac{1}{\Delta} & \frac1{\epsilon_2-\epsilon_j}
\end{array}
\right) \mathrm{det}{A^{\hat{1},\hat{2}}_{\hat{1},\hat{j}}}
\nonumber\\ &&+ \sum_{j=3}^{N-r} \epsilon^{2j}\ \mathrm{det}\left(\begin{array}{cc} 
-\frac{1}{\Delta} & \frac1{\epsilon_2-\epsilon_j} \\
 0 & \frac1{\epsilon_2-\epsilon_j} 
\end{array}
\right) \mathrm{det}{A^{\hat{1},\hat{2}}_{\hat{1},\hat{j}}}
\nonumber\\ &&+ \sum_{j=3}^{N-r}  \sum_{k>j}^{N-r}  \epsilon^{jk}\ \mathrm{det}\left(\begin{array}{cc} 
\frac1{\epsilon_2-\epsilon_j} & \frac1{\epsilon_2-\epsilon_k} \\
\frac1{\epsilon_2-\epsilon_j} & \frac1{\epsilon_2-\epsilon_k}
\end{array}
\right) \mathrm{det}{A^{\hat{1},\hat{2}}_{\hat{j},\hat{k}}}
\eea

\noindent where $A^{\hat{1},\hat{2}}_{\hat{j},\hat{k}}$ is the matrix $A$  from which line 1 and 2  as well as columns $j$ and $k$ have been removed while $\epsilon^{jk}$ is the sign of the corresponding partition of the indices. 

First, one easily sees that the two by two determinants $\mathrm{det}\left(\begin{array}{cc} 
\frac1{\epsilon_2-\epsilon_j} & \frac1{\epsilon_2-\epsilon_k} \\
\frac1{\epsilon_2-\epsilon_j} & \frac1{\epsilon_2-\epsilon_k}
\end{array}
\right) $ are all zero. Moreover, since a line permutation and the resulting minus sign makes $\mathrm{det}\left(\begin{array}{cc} 
0 & \frac1{\epsilon_2-\epsilon_j} \\
\frac{1}{\Delta} & \frac1{\epsilon_2-\epsilon_j}
\end{array}
\right) = \mathrm{det}\left(\begin{array}{cc} 
-\frac{1}{\Delta} & \frac1{\epsilon_2-\epsilon_j} \\
 0 & \frac1{\epsilon_2-\epsilon_j} 
\end{array}
\right)$, the change in the sign of the partition: $\epsilon^{1k} = - \epsilon^{2k}$, makes both single sums over $j$ also trivially cancel term by term. All in all, one simply has:
\bea
\lim_{\Delta \to 0}\mathrm{det} A = \frac{\mathrm{det} A^{\hat{1},\hat{2}}_{\hat{1},\hat{2}}}{\Delta^2}.
\eea 

 Consequently, the determinant has no simple pole at $\epsilon_1 \to \epsilon_2$; only a double pole $\frac{1}{(\epsilon_1-\epsilon_2)^2}$. Since the exact same reasoning can be carried out for any $\frac{1}{\epsilon_1 - \epsilon_j}$ ($j = 2,3,4 \dots N$), the resulting determinant can only depend on $\epsilon_1$ through double poles in all of its possible pairings. It can therefore be written as:

\bea
\mathrm{det} A = \sum_{j > 1} \frac{\mathrm{det} A^{\hat{1}, \hat{j}}_{\hat{1}, \hat{j}}}{\left(\epsilon_1 - \epsilon_j \right)^2} ,
\eea

\noindent since $\epsilon_1$ has to appear only twice in each term of the determinant (matrix elements coming from line and column 1), and the possible terms $ \frac1{\left(\epsilon_1 - \epsilon_j \right)^2}$ exhaust those two appearances of $\epsilon_1$.

This leads to a recursive construction of the determinant, since $ A^{\hat{1}, \hat{j}}_{\hat{1}, \hat{j}}$ has the exact same structure as the original matrix $M$ albeit with a dimension lowered by two and the absence of the indices $1$ and $j$. The exact same reasoning could then be applied to the possible pairings of the next unpaired index and the procedure can be kept going until $(N-r)/2$ pairs of indices have been made. The resulting determinant is then given by the sum (without sign alternances) over all the possible pairings.  Since $(-1)^{N-r}=1$, one therefore has:
\bea
K_\pi =\left(\prod_{k=1}^{N-r} 2 g F_x(\epsilon_{\bar{\pi}_k}) F_y(\epsilon_{\bar{\pi}_k})\right) 
 \sum_{\sigma \in \Sigma_{\bar{\pi}}} \prod_{k=1}^{(N-r)/2} \frac{1}{(\epsilon_{i_k} - \epsilon_{j_k} )^2}
 \label{kpi}
\eea

\noindent with the sum carried over $\Sigma_{\bar{\pi}}$, i.e.: the set of possible partitions in pairs $\sigma = \{\{i_1,j_1\},\{i_2,j_2\} \dots \{i_{(N-r)/2},j_{(N-r)/2}\} \}$ of the indices contained in $\bar{\pi}$.

The generic determinant operator $\hat{P}(r_1 \dots r_N)$, which become a projector when on-shell, has therefore been shown to have an explicit expansion in terms of only odd (for $N$ odd) or even (for $N$ even) products of $(r_i + \hat{R}_i)$. Moreover, the coefficient in front of each of these products is given by the sum, over possible pairings of the complementary indices, of the products of factors $\frac{1}{(\epsilon_i-\epsilon_j)^2}$, so that:
 
\bea 
\hat{P}(r_1 \dots r_N) &=&  \sum_{\pi \in \tilde{\Pi}} K_\pi \prod_{k=1}^r(r_{\pi_k} +R_{\pi_k})
\eea

\noindent where, for $N$ odd (even), $\tilde{\Pi}$ is the set of all subsets of $\{1,2 \dots N\}$ having exclusively odd (even) cardinality, while $K_\pi$ is given by eq. (\ref{kpi}). This result is to be contrasted with the square of the Pfaffian one would get from the determinant of a generic skew-symmetric matrix which would then be expanded a a double sum over these partitions into pairs.

As a simple explicit example for $N=4$ spins, the expansion of $\hat{P}(r_1, r_2, r_3, r_4)$ only contains $r=4, r=2$ and $r=0$ terms and is explicitly given by:

\bea
\hat{P}(r_1,r_2,r_3,r_4)  &=& (r_1+\hat{R}_1)(r_2+\hat{R}_2)(r_3+\hat{R}_3)(r_4+\hat{R}_4)
 \nonumber\\ &&+
\frac{4g^2F_x(\epsilon_3)F_y(\epsilon_3)F_x(\epsilon_4)F_y(\epsilon_4)}{(\epsilon_3-\epsilon_4)^2}
(r_1+\hat{R}_1)(r_2+\hat{R}_2)
 \nonumber\\ &&+
\frac{4g^2F_x(\epsilon_2)F_y(\epsilon_2)F_x(\epsilon_4)F_y(\epsilon_4)}{(\epsilon_2-\epsilon_4)^2}
(r_1+\hat{R}_1)(r_3+\hat{R}_3)
 \nonumber\\ &&+\frac{4g^2F_x(\epsilon_2)F_y(\epsilon_2)F_x(\epsilon_3)F_y(\epsilon_3)}{(\epsilon_2-\epsilon_3)^2}
(r_1+\hat{R}_1)(r_4+\hat{R}_4)
 \nonumber\\ &&+
\frac{4g^2F_x(\epsilon_1)F_y(\epsilon_1)F_x(\epsilon_4)F_y(\epsilon_4)}{(\epsilon_1-\epsilon_4)^2}
(r_2+\hat{R}_2)(r_3+\hat{R}_3)
 \nonumber\\ &&+
\frac{4g^2F_x(\epsilon_1)F_y(\epsilon_1)F_x(\epsilon_2)F_y(\epsilon_2)}{(\epsilon_1-\epsilon_2)^2}
(r_3+\hat{R}_3)(r_4+\hat{R}_4)  \nonumber\\ &&+ \left( \displaystyle \prod_{i=1}^4 2g F_x(\epsilon_i)F_y(\epsilon_i)\right) \cdot
\left[\frac{1}{(\epsilon_1-\epsilon_2)^2(\epsilon_3-\epsilon_4)^2}  \right.\nonumber\\ &&\left.\ \ \ +\frac{1}{(\epsilon_1-\epsilon_3)^2(\epsilon_2-\epsilon_4)^2}+\frac{1}{(\epsilon_1-\epsilon_4)^2(\epsilon_2-\epsilon_3)^2}\right].
\eea

While these expanded representations are certainly less compact than the determinant representation (\ref{opdet}) they reveal explicitly how the integrability of the system, and the resulting relations between $\Gamma$s come into play to in defining some of the fundamental underlying structure of the (eigen-)states constructed in this work.

\section{Conclusions}

In this article, we have showed how one can systematically build the eigenstates of spin-1/2 Richardson-Gaudin models in a magnetic field by only specifying the set of conserved charges and the eigenvalues associated with the given eigenstate. To do so, we defined a class of operators, parametrised by $N$ variables $(r_1 \dots r_N)$, which, when evaluated ``on-shell", i.e. at $(r_1 \dots r_N)$ solution to quadratic Bethe equations, explicitly become the projector on the corresponding normalised eigenstate.

This result completes the eigenvalue-based approach to these models by providing generic expressions for the eigenstates without having to define a Bethe Ansatz. One can hope that this construction could be used to provide expressions, written directly in terms of eigenvalues, for form factors of local operators which form the basic building blocks required to access the dynamical properties of these systems. 

\ack{The authors would like to thank Pieter W. Claeys for all of his valuable comments and suggestions on this work.}

\end{document}